\documentclass[preprint, superscriptaddress ,showpacs,preprintnumbers,amsmath,amssymb]{revtex4}
\usepackage{graphicx}
\usepackage[OT4]{fontenc}
\usepackage{dcolumn}
\usepackage{bm}

\begin{document}

\title{Background atoms and decoherence in optical lattices}

\author{Krzysztof Paw\l owski}
\affiliation{%
Center for Theoretical Physics, Polish Academy of Sciences, \\
Al. Lotnik\'ow 32/46, 02-668 Warsaw, Poland
}
\author{Kazimierz Rz\k{a}\.{z}ewski}%
\affiliation{%
Center for Theoretical Physics, Polish Academy of Sciences, \\
Al. Lotnik\'ow 32/46, 02-668 Warsaw, Poland
}
\affiliation{Faculty of Mathematics and Sciences, Cardinal Stefan Wyszy\'nski University, \\
ul. Dewajtis 5, 01-815, Warsaw, Poland}%

\begin{abstract}
All experiments with ultracold atoms are performed in the presence of background residual gas. With the help of a suitable master equation we investigate a role of these fast atoms on the loss of coherence in optical lattices. We present an exact solution of the master equation and give the analytic formulas for all correlation functions in the presence of one body losses. Additionally we discuss existing of a Schr\"odinger cat state predicted in this system in \cite{greiner2002}.
\end{abstract}


\maketitle

\section{Introduction}
Ultracold bosonic atoms placed in wells of the optical dipole potential formed by the interference of counter-propagating laser beams are studied in many recent experiments \cite{greiner2002qpt, duan2003, peredes2004, brennen1999}. A  Mott insulator state \cite{greiner2002qpt}, a collapse and revivals of the coherence in  the superfluid state \cite{greiner2002}, a massive entanglement between atoms \cite{mandel2003} and an Anderson localization \cite{roati2008} were all achieved. In a recent paper \cite{wuertz2009} a single site addressability has been demonstrated - a property essential for the future application of atoms in optical lattices in the quantum information processes. In all these processes a potentially significant role is played by losses. As  
always, the losses generate decoherence, a deterioration of the ability to interfere. In particular even the best achievable vacuum in the cell holding the ultracold atoms is never perfect. There are background atoms that are very fast. From time to time they collide  
with the relevant bosons that are practically at rest. As a result the atom trapped in the well is rapidly ejected and flies away. We are losing one atom but also the phase of the system undergoes a jolt. As a result, the atomic interference gets smeared-out.
In this paper we are studying in some detail this phenomenon. To this  
end we use a suitable master equation. We present an analytical  
solution for the evolution of a large class of correlation functions.  
Then, we use these solutions to discuss the first order coherence for  
the model of a perfect Mott insulator initial state and a model of a  
superfluid initial state. In the latter case we show that often used  
the coherent state as a representation of the state in a single well  
is sick. While correctly representing the distribution of the number  
of atoms found in a single site, it incorrectly breaks the $U(1)$  
symmetry of this state. We point out that discussed by some authors \cite{greiner2002, jack2003}  
Schr\"odinger cat state is an artifact of the coherent state  
representation.
\section{\label{sec:model}Model}

A motivation of our work is an experimental demonstration of a collapse and revivals of matter waves described in \cite{greiner2002}. The authors of this paper prepared bosonic $^{87}$Rb atoms in the superfluid state in the optical lattice. Suddenly they increased the intensity of lasers creating the deep optical lattice, such that the tunneling between lattice sites was completely suppressed. They hold the system during time $t$ and then switch-off the trapping potential, allowing cold atoms to expand. Next they take a picture of interfering matter waves. They were repeating the experiment extending only the holding time $t$. In the first few photos they recognized an interference pattern typical for a superfluid state \cite{greiner2002qpt}. Then this pattern had vanished just to a Gaussian noise, but at the time $t=t_{rev}$ it appeared again. We study the sensitivity of these revivals to decoherence due to non-perfect vacuum. Atoms from the residual thermal gas collide with ultracold atoms. We assume that the typical energy of an incoming atom from the background is so high that during the collision the condensed atom is kicked out from the lattice. This kind of one body losses leads inevitably to a reduced coherence in the system.

In order to describe the evolution of cold atoms in an optical lattice that includes these collisions we derived the following master equation:
\begin{equation}
\partial _t\hat{\rho }=-\frac{i}{\hbar }\left[\hat{H}_A, \hat{\rho }\right]+\gamma\sum _{k=1}^K  \left( \left[ \hat{a}_k, \hat{\rho } \hat{a}_ {k}^{\dagger }\right]+\left[ \hat{a}_k \hat{\rho}, \hat{a}_ {k}^{\dagger }\right] \right), 
\label{eqn:master}
\end{equation}
where $\hat{a}_ {k}$ annihilates one boson in the $k$th lattice site, $K$ is the number of lattice sites, $\gamma$ is a damping coefficient, and 
\begin{equation}
\hat{H}_A=\hbar\omega_0\sum_{k=1}^K \hat{a}_k^{\dagger}\hat{a}_k + \frac{U}{2}\sum_{k=1}^K\hat{a}_k^{\dagger}\hat{a}_k^{\dagger}\hat{a}_k\hat{a}_k ,
\label{eqn:ha}
\end{equation}
is the Bose-Hubbard Hamiltonian for atoms in an optical lattice in the absence of tunneling. The parameter $\hbar\omega_0$ in \eqref{eqn:ha} describes the kinetic and potential energy of one atom in the lattice, and $U$ is an average interaction energy per a pair of atoms. The details of the Hamiltonian $\hat{H}_A$ are given in \cite{jaksch1998}. We estimate the damping coefficient $\gamma$ and sketch the derivation of \eqref{eqn:master} in the appendix. Having analytical form of all coefficients in this model we are able to compare our theoretical results of this papers with the experimental papers. 

We assume that atoms in each well evolve independently so the final master equation is a simple extension of the master equation for a single well, examined more generally in \cite{anglin1997}. This type of master equation \eqref{eqn:master} was also a starting point in \cite{sinatra1998, jack2002, yun2008}.

In this paper we expand all states in the Fock basis, where $\mid n_1, \cdots, n_K\rangle$, means $n_1$ atoms in the first well, $n_2$ atoms in the second, etc. Here are three  interesting states:
\begin{itemize}
\item The superfluid state, called also multinomial state, with $N$ particles in $K$ potential wells
\begin{equation}
\mid \psi \rangle =\frac{1}{K^{N/2}}\sum _{n_1, \cdots, n_K} \sqrt{\binom {N}{n_1,\cdots, n_K}}\mid n_1, n_2, \cdots, n_K\rangle .
\label{eqn:multi}
\end{equation}
The sum in expression \eqref{eqn:multi} is over all non-negative integers, which adds-up to the initial number of atoms $N$, and $\binom {N}{n_1,\cdots, n_K} = \frac{N!}{n_1!\cdots n_K!}$ is the multinomial symbol. This state is in fact a $SU(K)$ coherent state \cite{buonsante2005}.
\item The Mott insulator state, which describes the system with exactly $n$ atoms in each well:
\begin{equation}
\mid \psi \rangle =\bigotimes_{k=1}^K\mid n\rangle .
\label{eqn:mott}
\end{equation}
\item The coherent state 
\begin{equation}
\mid \psi \rangle  = \bigotimes_{k=1}^ {K} \left(\sum _{n=0}^{\infty} \frac{\alpha^n e^{-\left|\alpha\right| ^2/2}}{\sqrt{n!}}\mid n\rangle\right) .
\label{eqn:coh}
\end{equation}
Although the coherent state violates the barionic superselection rule, it seems to work quite well as an approximation of the superfluid with the total number of atoms $N=K\left| \alpha\right|^2$. The evolution of this state in a context of optical lattices was considered i.e. in \cite{greiner2002, jack2002}.
\end{itemize}

\section{\label{sec:generating}Generating function}
The master equation \eqref{eqn:master} describes a system of $K$ independent modes, with the interaction in each mode and with the one body losses. In the next section \ref{sec:decoherence} we use this model to describe the influence of background atoms for coherence properties in optical lattices. We stress that equation \eqref{eqn:master} has quite general form and might be appropriate for other systems. 
In this section we present the exact solution of a master equation of this type. 

Let us consider a dynamics of an average value of the normal ordered operator:
\begin{equation}
\left\langle \prod_{k=1}^K\left(\hat{a}_ {k}^{\dagger}\right)^{r_k}\prod_{k=1}^ K\hat{a}_ {k}^{l_k}\right\rangle .
\label{eqn:cor_wells}
\end{equation}

At this point we assume for simplicity that for all indices $k$, coefficient $r_k$ or $l_k$ is equal to $0$. It means $r_k+l_k=max (r_k, l_k)$.

We focus on the most important case of expression \eqref{eqn:cor_wells}, correlation functions, where the number of creation operators and annihilation operators is the same. It means $\sum_{k=1}^K l_k=\sum_{k=1}^K r_k = R$, where $R$ is the order of correlation function. Below we will often use this property. 

The density operator has the following matrix form in the Fock basis:
\begin{equation}
\hat{\rho}=\sum_{\substack{n_1, \cdots, n_K\\
			m_1, \cdots, m_K }}^N \rho^{n_1,\cdots, n_K}_{m_1,\cdots, m_K} \mid n_1, \cdots, n_K\rangle\langle m_1,\cdots, m_K\mid.
\label{eqn:denisty}
\end{equation}
The density matrix elements which are appropriate to calculate \eqref{eqn:cor_wells} are of the form $\rho^{n_1+r_1,\cdots,n_K+r_K}_{n_1+l_1,\cdots, n_K+l_K }$, where $n_k$ are non-negative integer numbers. We introduce a shorter notation for indices $\{n_1+l_1,\cdots, n_K+l_K\} =\bm{n}+\bm{l} $. Thus $\bm{l}$ belongs to $\mathbb{N}^K$ and with $\bm{\varepsilon}_k$ we denote the versor in the $k$th direction in this space ($\left(\bm{\varepsilon}_k\right)_i=\delta_{k, i} $).
 From the master equation \eqref{eqn:master} we know their evolutions
\begin{eqnarray}
\nonumber\partial _t\rho^{\bm{n}+\bm{r}}_{\bm{n}+\bm{l}}&=\dot{\imath}\lambda \sum_k^K(r_k-l_k)\left( 2n_k + r_k+l_k-1\right)\rho^{\bm{n}+\bm{r}}_{\bm{n}+\bm{l}}+\\
\nonumber & +2\gamma \sum_k^K \sqrt{(n_k+1) (n_k+r_k+l_k+1)}\rho^{\bm{n}+\bm{r}+\bm{\varepsilon}_k}_{\bm{n}+\bm{l}+\bm{\varepsilon}_k}\\
& -\gamma \sum_k^K(2n_k+r_k+l_k)\rho^{\bm{n}+\bm{r}}_{\bm{n}+\bm{l}},
\label{eqn:denscoef}
\end{eqnarray}
where $\lambda=U/(2\hbar)$.

We define the generating function $h^{\bm{r}}_{\bm{l}}$:
\begin{equation}
h^{\bm{r}}_{\bm{l}}\left(\bm{x}; t\right)=\sum_{\bm{n}} x_1^{n_1} \ldots x_K^{n_K} \rho^{\bm{n}+\bm{r}}_{\bm{n}+\bm{l}} \prod_{k=1}^K\prod_{m=1}^{r_k+l_k}\sqrt{n_k+m} ,
\label{eqn:fun_wells}
\end{equation}
where $\bm{x}$ is any vector in $\mathbb{R}^K$ space.
The function $h^{\bm{r}}_{\bm{l}}$ in \eqref{eqn:fun_wells} is defined in such a way that along a line $x_1=\cdots=x_K=1$ is just equal to \eqref{eqn:cor_wells}. Multiplying the equation \eqref{eqn:denscoef} by $\prod_{k=1}^Kx_k^{n_k}\prod_{m=1}^{r_k+l_k}\sqrt{n_k+m}$ and then taking the sum over all indices $n_k$ we derive an equation for $h^{\bm{r}}_{\bm{l}}$
\begin{eqnarray*}
\partial _th^{\bm{r}}_{\bm{l}}&=&\sum_{k=1}^K\left(2\gamma -\left(2\gamma -2\dot{\imath}\lambda\left(r_k-l_k\right)\right)x_k\right)\partial _{x_k} h^{\bm{r}}_{\bm{l}}+\\
& & +\sum_{k=1}^K\left(\dot{\imath}\lambda \left(r_k^2-l_k^2\right)-2\gamma R\right)h^{\bm{r}}_{\bm{l}} .
\end{eqnarray*}
This is the first order partial differential equation, so one can find the solution by the method of characteristics:
\begin{eqnarray}
\nonumber h^{\bm{r}}_{\bm{l}}\left(x_1, \cdots, x_K, t\right)=e^{\sum_{k=1}^K\left(\dot{\imath} \lambda \left(r_k^2-l_k^2\right)-2\gamma R\right)t}\times\\
\times h^{\bm{r}}_{\bm{l}}\left(f\left(x_1, r_1-l_1, t\right), \cdots, f\left(x_K, r_K-l_K, t\right), 0\right),
\label{eqn:solution}
\end{eqnarray}
where $f\left(x, r, t\right)=\frac{2 \gamma - \left( 2\gamma -\kappa_r x\right) e^{-\kappa_r t} }{\kappa_r} $ and $\kappa_r = 2 \gamma - 2 \dot{\imath} \lambda r$.

From the generating function one can easily extract all density matrix terms:
\begin{equation}
\rho^{\bm{n}+\bm{r}}_{\bm{n}+\bm{l}}=\frac{\partial_{x_1}^{n_1}\ldots\partial_{x_K}^{n_K}h^{\bm{r}}_{\bm{l}}\left(0,\cdots, 0;t\right)}{\prod_{k=1}^K \sqrt{\left(n_k+l_k\right) !\left(n_k+r_k\right)!}} 
\label{eqn:all_density}
\end{equation}
Solutions \eqref{eqn:solution} and \eqref{eqn:all_density} are valid for all initial states and any number of wells. They cover exact solutions of papers \cite{radka2004, sinatra1998, yun2009} as special cases. The paper  \cite{radka2004} treats all correlation functions in optical lattices but without losses. The papers \cite{sinatra1998, yun2009} are interesting discussions of one, two and three body losses in two mode system. There are presented exact solution for low order correlation function in the case of one body losses using the stochastic wave approximation.

The equation \eqref{eqn:all_density} says that the terms of density matrix decay exponentially with the rate growing as we go away from the diagonal.  For long times all off-diagonal terms vanish and the state of the system becomes completely mixed, in agreement with the general properties of the decoherence.

\section{\label{sec:decoherence}Decoherence}
In this section we will use the general solution \eqref{eqn:all_density} for the three states introduced in the section \ref{sec:generating}. Subscripts 'coh', 'sf' and 'mi' denote the coherent state, the superfluid state and the Mott insulator, respectively. For shorter notation we omit indices $\bm{r}$ and $\bm{l}$, namely $h^{\bm{r}}_{\bm{l}}\equiv h$. The appropriate generating functions have forms
\begin{eqnarray}
\nonumber h_{sf}\left(\bm{x}; t\right)&=&\frac{\prod_{j=0}^{R-1}\left(N-j\right)}{K^N} e^{-\gamma R t+\dot{\imath}\lambda\sum_{k=1}^K \left(r_k^2-l_k^2 \right) t}\times\\
& &\times\left( \sum_{k=1}^ {K}f\left(x_k, r_k-l_k, t\right)\right)^{N-R}\label{eqn:solsf}\\
h_{coh}\left(\bm{x}; t\right)&=&\prod_{k=1}^K \left(\alpha^{\ast}\right)^{r_k}\left(\alpha\right)^{l_k}e^{\left| \alpha\right|^2 \left(f\left(x_k, r_k-l_k, t\right)-1\right)} \label{eqn:solcoh}\\
h_{mi}\left(\bm{x}; t\right)&=&\prod_{k=1}^K \left(1-(1-x_k e^{-2\gamma t})\right)^{N/K} \label{eqn:solmi}
\end{eqnarray}
Using these functions we can analyze all correlation functions for these three different initial states. It is not surprising that for all three states the average number of atoms per well is given by the same formula
\begin{equation}
\langle \hat{a}_k^{\dagger}\hat{a}_k\rangle =\partial_{x_k} h_{\bm{0}}^{\bm{0}}\left( 1, \cdots, 1;t\right)=\frac {N} {K}e^{-2\gamma t} .
\end{equation}
This exponential decay of the particle number was observed in several experiments. The coefficient $(2\gamma)^{-1}$ is known in literature as the lifetime of the trap. In the appendix we give the straightforward but lengthy derivations of $\gamma$:
\[
\gamma=\sigma_{cross} \rho_b \langle v \rangle /\sqrt{6}, 
\]
where $\sigma_{cross}$ is a cross section for the collision between 'hot' and cold atoms, $\rho_b$ is a density of the background atoms and $ \langle v \rangle$ is their average velocity.
We compare measured values of $\gamma$ with our analytical estimation getting quantitative agreement \cite{gerton1999, myatt1997, mewes1996}, except the experiments with the metastable helium $^{\ast}$He \cite{bardou1992, browaeys2000}. We suppose that in experiments with helium the interaction with background atoms is not the main source of one body losses.
\subsection{First order correlation function}
More interesting is the evolution of the first order correlation function
\[ g^{1} = \frac{\langle\hat{a}_i^{\dagger}\hat{a}_j\rangle}{\sqrt{\langle\hat{a}_i^{\dagger}\hat{a}_i\rangle\langle\hat{a}_j^{\dagger}\hat{a}_j\rangle}} ,\]
which in the experiment of Bloch \cite{greiner2002}, is responsible for the revivals and collapses of the interference pattern. For the Mott insulator state this function is just equal to $0$ -- to describe interference of matter waves in this state higher order correlation function are needed. If the initial state is a superfluid or a coherent state one get
\begin{eqnarray}
g_{sf}^{1}&=&\left(1+\frac{w(t)} {K} \right)^{N-1}\\
g_{coh}^{1}&=& e^{\left|\alpha\right| ^2 w(t)},
\end{eqnarray}
where
$w(t)= 2 \left(\frac{ \gamma^2 + \left(\gamma \lambda \sin (2\lambda t) + \lambda^2 \sin (2\lambda t)\right) e^{-2\gamma t}}{\gamma^2+\lambda^2} -1\right)$.
In the limit of large number of wells and large number of atoms, both first order correlation functions are identical and equal to $g^{1}=1+\left|\alpha\right| ^2 w(t)$. Thus, in typical experiments with optical lattice the damping of coherence is the same for superfluid and coherent state. It suggests that the coherent state can be used instead of multinomial one also for studying the decoherence in optical lattice, but in the next section we are showing the limitation of this replacement.

\begin{figure}
\includegraphics[width= 230pt, angle=0, clip=true]{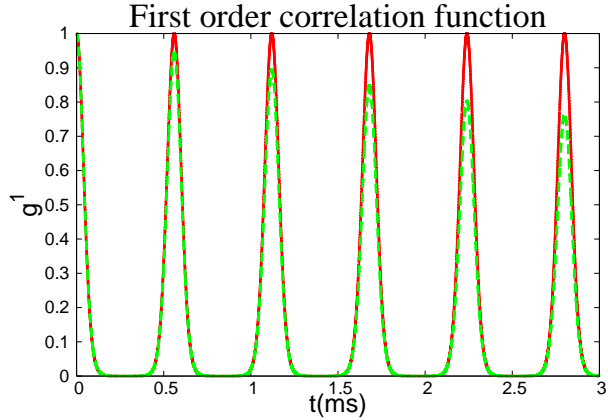}
\caption{ (Color online)The first order correlation function for systems initially in superfluid or coherent state. Solid red line, -- the damping like in recent experiments $\gamma = 0.001$/s, green dashed line -- with  much stronger damping $\gamma = 10.0$/s. Number of wells $8000$ and number of atoms $20000$ -- like in the experiment \cite{greiner2002}.}
2\label{fig:first}
\end{figure}
We present these functions in figure \ref{fig:first}.
Indeed the evolution of the first order correlation function has a structure of collapses and revivals at the multiples of the revival time $t_{rev}=\frac{\pi}{\lambda}$. For experimental value of parameter $\gamma$ no effects of decoherence can be seen. It means that in the experiment \cite{greiner2002} there are more important sources of decoherence. As authors of this paper suggest it can be the noise of the laser beams. 

\begin{figure}
\begin{minipage}[t]{0.45\textwidth}
\centering
\includegraphics[width= \textwidth, angle=0, clip=true]{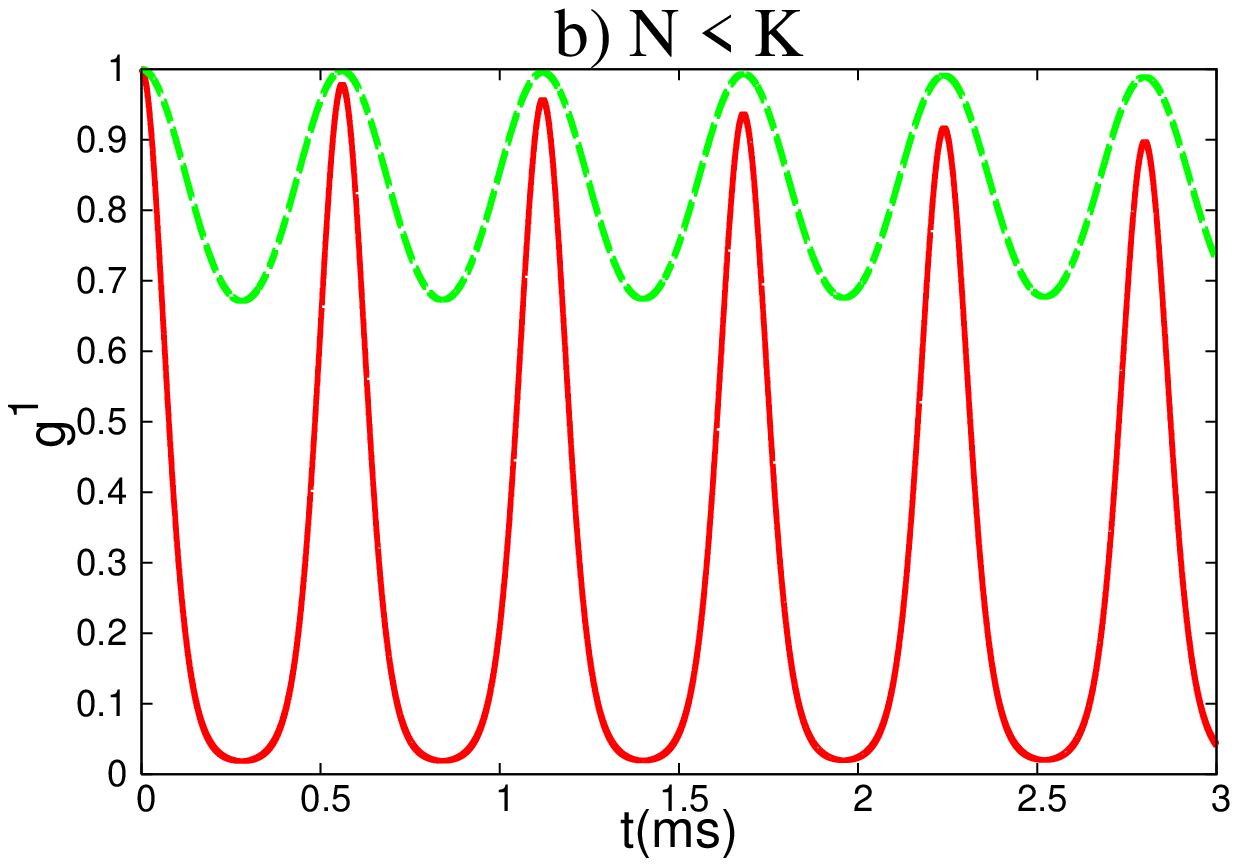}
\end{minipage}
\hspace{0.09\textwidth}
\begin{minipage}[t]{0.45\textwidth}
\centering 
\includegraphics[width= \textwidth, angle=0, clip=true]{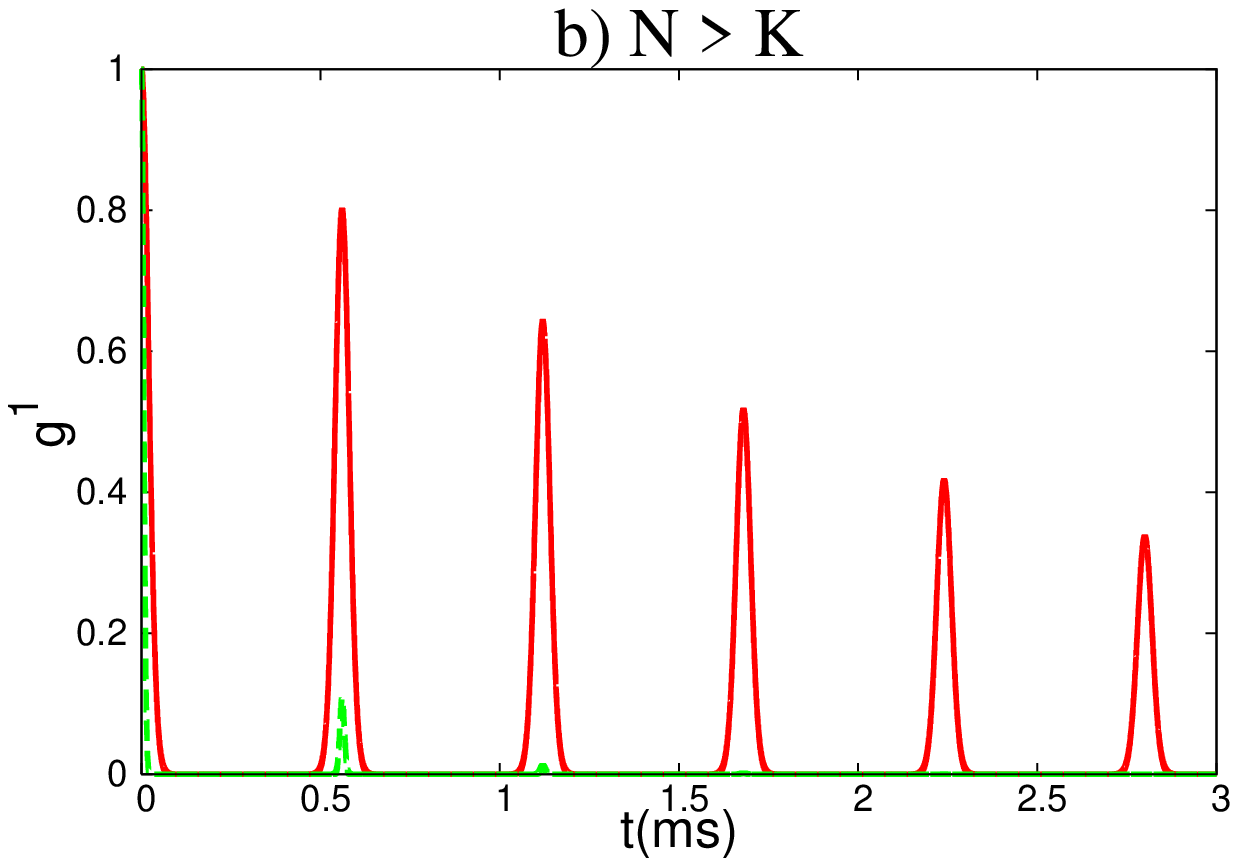}
\end{minipage}
\caption{ (Color online)Two different regimes for evolution first order correlation function: a) number of atoms $1000$ and wells: solid red line $1000$, dashed green line $10000$. b)number of wells $1000$ and atoms: green line $10000$, red line $100000$. Damping coefficient in all cases $\gamma=10/s$.}.
\label{fig:regimes}
\end{figure}

We investigate the decoherence varying the experimental parameters - the number of atoms, the number of potential wells and the damping parameter $\gamma$, but holding constant the interaction parameter $U$. 
We observe that for bigger ratio $K/N$ revivals are broader and collapses of coherence are not complete (see figure \ref{fig:regimes}a).
In it case even for large damping, the function $g^1$ tends nor to $0$, but to some positive value. In the case of strong damping it is difficult to determine if the master equation is still valid -- it should be verified experimentally.
On the other hand if we increase the total number of atoms, the revival becomes extremely narrow and extremely hard to detect. It is worth to stress again, that we hold the parameter $U$ constant and equal to the value from the experiment \cite{greiner2002}. In fact $U$ can be changed so the time scale might be totally different (like in typical experiments with a double well potential). 

We present this two regimes in the figure \ref{fig:regimes}. One can easily see that the first regime, where the number of potential wells is bigger than the number of atoms is less susceptible to decoherence. Increasing the number of atoms leads to a fast decay of coherence.

%
\section{\label{sec:qfncts}Q-functions -- death of a Schr\"odinger cat}
Up to now we haven't seen any notable consequences of using coherent states for the superfluid state. Despite the coherent state in the context of matter wave violates the fundamental law of nature, the barionic number superselection, it seems to be indistinguishable from the superfluid state. 
In the kind of thermodynamic limit the distribution of number of atoms per one well is identical, the damping of number of particle is the same, and the first and the second order correlation functions are so similar in both cases that in any experiment they cannot be distinguished \cite{radka2004}. 
On the other hand the coherent state is much easier to handle than the multinomial one. Hence it is no wonder that instead of multinomial state usually only its coherent counterpart is investigated. In its evolution at the time $\frac{1}{2} t_{rev}$ surprisingly appears a Schr\"odinger catlike state --  the state of the system is a superposition of two coherent states with opposite phases \cite{greiner2002}. It was even predicted in \cite{jack2003} that this superposition state may stable even in the presence of one and three body losses. To shed some light on this issue we compute one of quasiproababilty functions. The simplest here is the Husimi Q-function:
\begin{displaymath}
Q_{\hat{\rho}}\left(\bm{\beta}\right)=\frac{1}{\pi^K}\langle \bm{\beta}\mid\hat{\rho}\mid\bm{\beta}\rangle ,
\end{displaymath}
where $\bm{\beta}$ is a vector in $K$ dimensional complex space $\mathbb{C}^K$ and $\mid \bm{\beta}\rangle=\bigotimes_{k=1}^K\mid \beta_k\rangle$ is a product of coherent states in each well.

We use the generating functions \eqref{eqn:solcoh}, \eqref{eqn:solmi} and \eqref{eqn:solsf} and the general form for the denisty matrix \eqref{eqn:all_density} to calculate the Husimi functions for our three initial states in the simplest $\gamma=0$ case:
\begin{eqnarray*}
Q_{coh}\left(\bm{\beta};t\right) &=&\frac{1}{\pi^K}\prod_{k=1}^K\sum_{n, m}^{\infty}\frac{\left(\alpha\beta_k\right)^n \left(\alpha^{\ast}\beta_k^{\ast}\right)^m}{n!m!}\times \\
& &\times e^{\left|\alpha\right|^2-\mid\beta_k\mid^2}e^{\dot{\imath} \lambda t \left(m-n\right)\left(m+n-1\right)} ,\\
Q_{sf}\left(\bm{\beta};t\right) &=&\frac{1}{\pi^K K^N N!}\left|\sum_{\bm{n}}^{N} \binom {N}{n_1, \cdots, n_k} e^{\dot{\imath} \lambda t n_k^2} \prod_{k=1}^K\beta_k^{n_k}\right| ^2 ,\\
Q_{mi}\left(\bm{\beta};t\right) &=& \frac{1}{\pi^K}\prod_{k=1}^K\frac{e^{-\left|\beta_k\right|^{2N/K}} \left|\beta_k\right|^{2 n}}{(N/K)!} .
\end{eqnarray*}

\begin{figure}
\includegraphics[width=1.0\textwidth, clip]{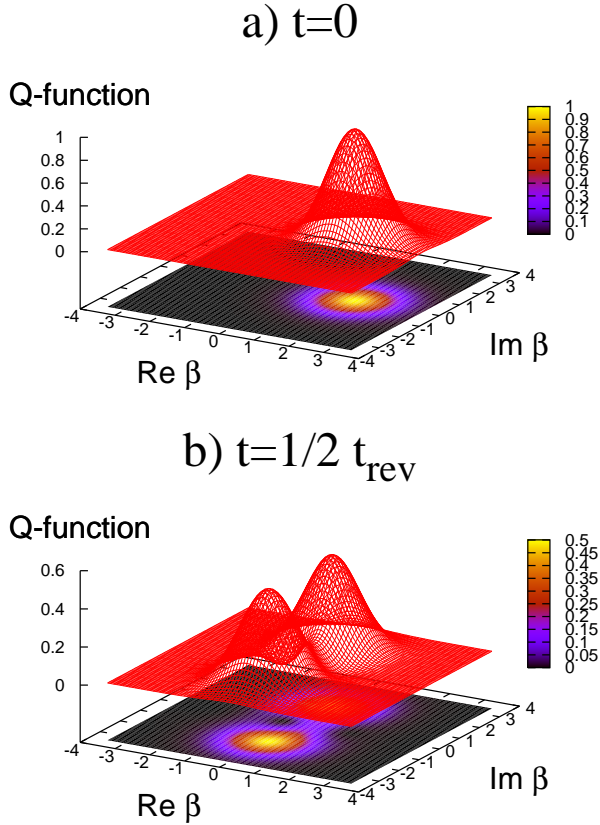}
\caption{ (Color online)The Husimi function $Q(t)$ at times a) $0$, b) $\frac{1}{2}t_{rev}$ if the initial state is coherent. At the time $\frac{1}{2}t_{rev}$ the system is a superposition of two coherent states.}
\label{fig:qcoh}
\end{figure}

\begin{figure}
\includegraphics[width=1.0\textwidth, clip]{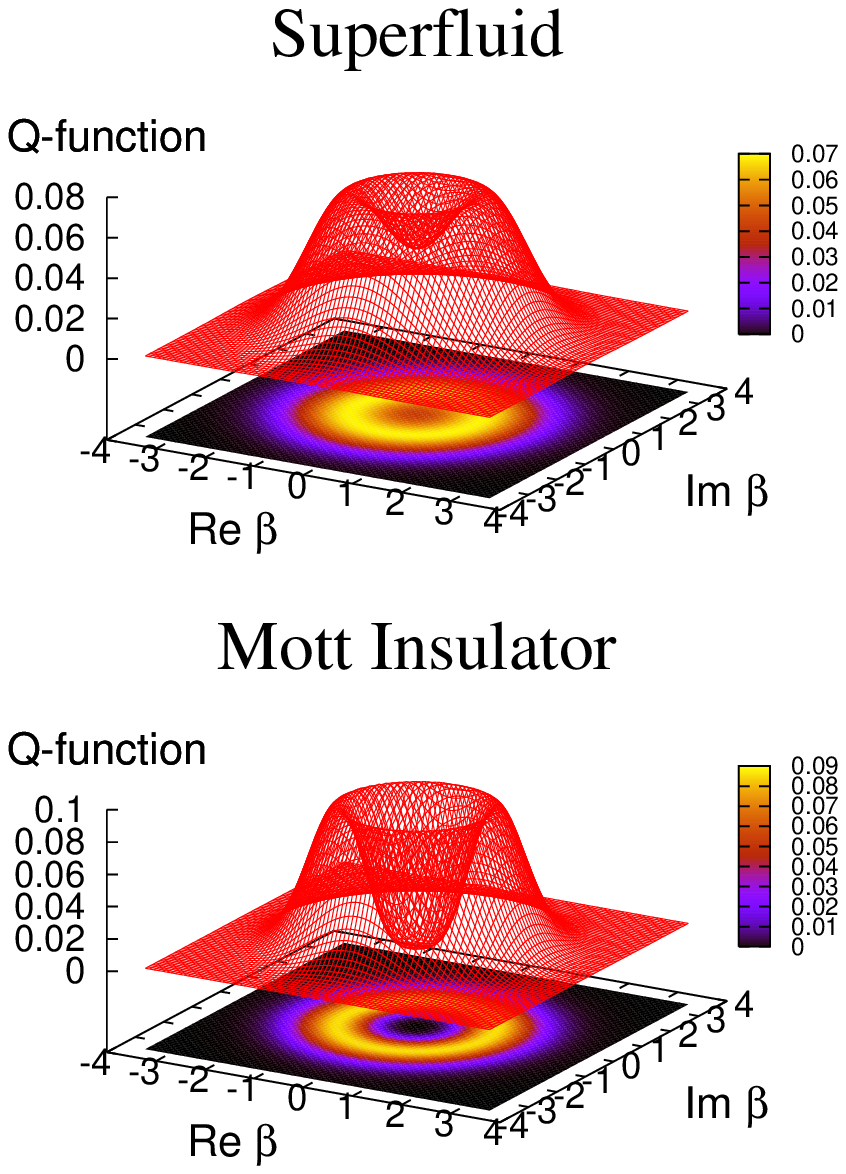}
\caption{ (Color online)The Husimi functions for a) the Mott insulator state b) the superfluid state. The Husimi function in both cases is time independent and has U(1) symmetry.}
\label{fig:qsf}
\end{figure}
The immediate conclusion is that the Q-functions for the coherent and the superfluid state are quite different. The first one at time $t=0$ is a product of Gaussian functions around points $\alpha$, while the second function depends only on the absolute value of $\alpha$, so it is symmetric around $0$. To visualize the Husimi functions, we integrate them over $(K-1)$ lattice sites
\[ Q\left(\beta ;t  \right)=\int \mbox{d} \beta_2 \cdots\mbox{d}\beta_K\ Q\left(\beta, \beta_2, \cdots,\beta_K ;t\right) \]
These reduced Husimi functions are presented in figures \ref{fig:qcoh} and \ref{fig:qsf}. The reduced Husimi function of the superfluid and the Mott insulator state is time independent. 

Thus, we conclude that the much discussed Schr\"odinger cat for atoms in the optical lattice is an artifact of the coherent state representation of the superfluid state. Our conclusion would be the same if we used the Wigner function rather than the Husimi function.

\section{\label{sec:concl}Conclusions}

The most important result of this paper is the exact solution for the master equation \eqref{eqn:master}. 
We use this solution to discuss the influence of 'hot' background atoms for the coherence properties of ultracold atoms in optical lattices. 
The calculated damping of the average number of particles agrees with experiments. In the main experiment under consideration \cite{greiner2002} the damping of $g^1$ due to collisions with the background atoms in negligible. 
We consider additional isotropic harmonic confinement also. Then in each potential well atoms have an additional energy shift which value depend on the position of the site. The spread of the shift's values is however so small, that the effects related to them are negligible. 
Studying further this model we find interesting regimes in which the function $g^1$ does not collapse, even for large losses. It means that the interference pattern should be seen in this regime at any time so the decoherence is quenched. Furthermore we have shown that using a coherent state for description of cold atoms might lead to wrong conclusions. The Schr\"odinger cat is an example.

\begin{acknowledgements}
The authors acknowledge financial support of the Polish Geovernment Research Funds for 2009-2011.
\end{acknowledgements}

\appendix*
\section{\label{appendix1}Coefficient $\gamma$}
We present an estimation of the coefficient $\gamma$ in the master equation \eqref{eqn:master}. To calculate this parameter one must follow whole derivation of the master equation. The origin of $\gamma$ lies in the second term in the Hamiltonian:
\begin{widetext}
\begin{equation}
\hat{H}=\int \mbox{d}^3 r \hat{\psi}^{\dagger}\left(  \bm{r}\right) \mathcal{H}_0 \hat{\psi}\left(  \bm{r}\right)  + \frac{1}{2}\int \mbox{d}^3r\mbox{d}^3r^{\prime} \hat{\psi}^{\dagger}\left(  \bm{r}\right) \hat{\psi}^{\dagger}\left( {\bm{r}^{\prime}}\right) V\left( \bm{r}- \bm{r}^{\prime} \right) \hat{\psi}\left( {\bm{r}^{\prime}}\right)\hat{\psi}\left(  \bm{r}\right) + \cdots,
\label{eqn:full}
\end{equation}
where an operator $\hat{\psi}^{\dagger}\left(  \bm{r}\right)$ creates a bosonic field,  $\mathcal{H}_0=-\frac{\hbar^2 \nabla^2}{2 m}+V (\bm{r})$ consists of the kinetic energy and a periodic potential and $V\left( \bm{r}- \bm{r}^{\prime} \right)$ is a potential of interaction between particles.
We represent the field operator $\hat{\psi}\left(  \bm{r}\right)$ as a superposition of BEC and background gas modes
\begin{equation}
\hat{\psi}^{\dagger}\left(  \bm{r}\right) = \phi \left( \bm{r}\right) \hat{a} + \frac{1}{\sqrt{V}}\sum_{  {k} } e^{\dot{\imath}  \bm{k}  \bm{r} }\left( \hat{c}_{  {k} }+\hat{d}_{  {k} } \right),
\label{eqn:fieldop}
\end{equation}
where $\hat{a}$ annihilates one boson in a ground state, $\phi \left( \bm{r}\right)$ is a spatial representation of the ground state fo $\mathcal{H}_0$, $\hat{c}_{  {k} }$  annihilates one background atom with wave-vector $\bm{k}$ and $\hat{d}_{\bm{k} }$ annihilates one boson with wave-vector $ \bm{k}$, which has been kicked out from the condensate. All non-condensed bosons are modeled by the plane waves in quantization volume $V$. We introduce two annihilation bosonic operator $\hat{d}_k$ and $\hat{c}_k$, because in general, ultracold atoms and background gas might be formed from different species. 

At this point we make the crudest simplification, we replace $V\left( \bm{r}- \bm{r}^{\prime} \right)$ with the contact pseudo-potential $\frac{4\pi\hbar^2 a}{m}\delta\left( \bm{r}- \bm{r}^{\prime}\right)$, where $a$ is th s-wave scattering length, and $m$ is a reduced mass of colliding particles. This approximation works quite well for collision of two ultracold atoms and it is not justified for a collision between one cold and one 'hot' background atom. Still, this simplification offers some insight into this process. 
We expect that the final result is at least of the same order of magnitude as the accurate damping constant $\gamma$. 

We limit our consideration only to the first two terms in \eqref{eqn:full}. Substitution \eqref{eqn:fieldop} into \eqref{eqn:full} leads to plenty of terms. Most of them have no influence on the values of $\gamma$. The significant part is given by the Bose-Hubbard Hamiltonian 
\begin{displaymath}
\hat{H}=\hat{H}_A+\hat{H}_{env}+\hat{H}_{int}.
\end{displaymath}
It consists of three terms.
\begin{itemize}
\item The free evolution of ultracold gas in an optical lattice.

$\hat{H}_A=\hbar\omega_0\hat{a}^{\dagger}\hat{a} + \frac{U}{2}\hat{a}^{\dagger}\hat{a}^{\dagger}\hat{a}\hat{a} ,$

where $\hbar\omega_0$ is the energy of a single atom in an optical lattice, $U$ is an interaction coefficient

\item The term which describes the evolution of free 'hot' atoms
$\hat{H}_{env}=\hbar\sum_{\bm{k}}\omega_k\left(\hat{c}_ {\bm{k}}^{\dagger}\hat{c}_ {\bm{k}} + \hat{d}_ {\bm{k}}^{\dagger}\hat{d}_ {\bm{k}}\right),$
where  $\hbar\omega_{k}=\frac{\hbar^2 \bm{k}^2}{2m}$ is a kinetic energy of a free particle of mass $m$ and momentum $\hbar  \bm{k}$. 


\item The interaction between cold and 'hot' atoms
\begin{eqnarray}
\nonumber H_{int}&=&\frac{g}{2V^{3/2}}\int \mbox{d}^3r\sum_{ \bm{k}_1,  \bm{k}_2,  \bm{k}_3}\hat{c}_{ \bm{k}_1}^{\dagger}\hat{d}_{\bm{k}_2}^{\dagger}\hat{c}_{ \bm{k}_3}\hat{a}\,\phi\left(  \bm{r}\right)e^{\dot{\imath}\left( \bm{k}_1- \bm{k}_2- \bm{k}_3\right)  \bm{r}} +h.c. =\\
& & =\hat{\gamma}_{ \bm{k}_1,  \bm{k}_2,  \bm{k}_3}\hat{c}_{ \bm{k}_1}^{\dagger}\hat{d}_{ \bm{k}_2}^{\dagger}\hat{c}_{ \bm{k}_3}\hat{a}+h.c. ,
\label{eqn:notation}
\end{eqnarray}
where $g=\frac{4\pi\hbar^2 a}{m}$. For shorter notation we have introduced in \eqref{eqn:notation} the operator $\hat{\gamma}_{ \bm{k}_1,  \bm{k}_2,  \bm{k}_3}$.
\end{itemize}
We follow the derivation of master equation based on \cite{carmichael1991}. Our derivation is simplified in comparison with \cite{anglin1997}. We quote it here to obtain the expression for the damping coefficient $\gamma$.
We write the von Neumann equation in the interaction picture
\begin{equation*}
\frac{d\tilde{\rho}}{dt}=-\frac{\dot{\imath}}{\hbar}\left[\tilde{H}_{int}\left(t\right),\;\tilde{\rho} \right],
\end{equation*}
where with a tilde we denote the transformation
\begin{displaymath}
 \tilde{O} =  e^{\dot{\imath}\left( \hat{H}_{A} + \hat{H}_{E}\right)t/\hbar}\hat{O} e^{-\dot{\imath} \left( \hat{H}_{A} + \hat{H}_{E}\right)t/\hbar}
\end{displaymath}
Using the von Neumann equation one can easily write the first two terms of perturbation series of the formal solution of the density matrix (Born approximation):
\begin{equation}
\frac{d\tilde{\rho}}{dt}=-\frac{\dot{\imath}}{\hbar}\left[\tilde{H}_{int}\left(t\right),\;\tilde{\rho} \left( 0\right) \right]-\frac{1}{\hbar^{2}}\int _{0} ^{t}\mbox{d}\tau\,\left[\tilde{H}_{int}\left(t\right) ,\left[\tilde{H}_{int}\left(\tau\right), \tilde{\rho}\left(\tau\right)\right]\right]
\label{eqn:formalne} 
\end{equation}
In the next steps we trace both sides over the environment and transform it back to the Schr\"odinger picture. We assume that initially the density operator of the system has a form 
\begin{displaymath}
\hat{\rho}=\hat{\rho}_{BEC}\otimes\hat{\rho}_E,
\end{displaymath}
where $\hat{\rho}_{BEC} $ is a density operator of BEC and $\hat{\rho}_E$ is a density operator of the environment. In the derivation we also assume that the change of $\tilde{\rho}_{BEC}$ during time $t$ is negligible and the evolution is well approximated by the Markov process. 
All this operation one can find in the following formula for coefficient $\gamma$:
\begin{eqnarray}
\gamma \hat{a}\hat{\rho}_{BEC}\hat{a}^{\dagger} = -\frac{1}{\hbar^2} \int d\tau\hat{\gamma}_{ \bm{k}_1,  \bm{k}_2,  \bm{k}_3} \hat{\gamma}_{ \bm{k}_1^{\prime},  \bm{k}_2^{\prime},  \bm{k}_3^{\prime}} \hat{a}\hat{\rho}_{BEC}
\hat{a}^{\dagger} e^{\dot{\imath} \left(\tau-t\right)\Delta} Tr\left\{\hat{c}_{ \bm{k}_1}^{\dagger}\hat{c}_{ \bm{k}_3}\hat{d}_{ \bm{k}_2}
\hat{\rho}_E
\hat{d}_{ \bm{k}_2^{\prime}}^{\dagger}\hat{c}_{ \bm{k}_3^{\prime}}^{\dagger}\hat{c}_{ \bm{k}_1^{\prime}}\right\},
\label{eqn:gamma_gen}
\end{eqnarray}
where $\Delta=U/\hbar\,\,\hat{a}^{\dagger}\hat{a} + \omega_0+\omega_{\bm{k}_1} - \omega_{\bm{k}_2}-\omega_{\bm{k}_3}$.

We assume that the background atoms are in a thermal state and the atoms from $d$-modes are initially in the vacuum state $\left| 0\rangle\langle 0\right|$. Thus 

$Tr\left\{\hat{c}_{ \bm{k}_1}^{\dagger}\hat{c}_{ \bm{k}_3}\hat{d}_{ \bm{k}_2}
\hat{\rho}_E
\hat{d}_{ \bm{k}_2^{\prime}}^{\dagger}\hat{c}_{ \bm{k}_3^{\prime}}^{\dagger}\hat{c}_{ \bm{k}_1^{\prime}}\right\}  = \langle \hat{c}_{ \bm{k}_1}^{\dagger}\hat{c}_{ \bm{k}_1}\rangle \left( 1+ \langle \hat{c}_{ \bm{k}_3}^{\dagger}\hat{c}_{ \bm{k}_3}\rangle\right) \delta_{ \bm{k}_1}^{ \bm{k}_1^{\prime}} \delta_{ \bm{k}_2}^{ \bm{k}_2^{\prime}} \delta_{ \bm{k}_3}^{ \bm{k}_3^{\prime}}$

For high temperature the average occupation of any energy level, $\langle
 \hat{c}_{\bm{k}}^{\dagger}\hat{c}_{\bm{k}} \rangle$, is much less than unity what justifies the approximation:
\begin{displaymath}
\langle \hat{c}_{ \bm{k}_1}^{\dagger}\hat{c}_{ \bm{k}_1}\rangle \langle 1+\hat{c}_{ \bm{k}_3}^{\dagger}\hat{c}_{ \bm{k}_3}\rangle \approx \langle \hat{c}_{ \bm{k}_1}^{\dagger}\hat{c}_{ \bm{k}_1}\rangle
\end{displaymath}
After that the formula for $\gamma$ is reduced to:
\begin{equation}
\gamma \hat{a}\hat{\rho}_{BEC}\hat{a}^{\dagger} = -\frac{1}{\hbar^2} \hat{\gamma}_{ \bm{k}_1,  \bm{k}_2, \bm{k}_3}^2 \hat{a}\hat{\rho}_{BEC}
\hat{a}^{\dagger} e^{\dot{\imath} \left(\tau-t\right)\Delta} \langle \hat{c}_{ \bm{k}_1}^{\dagger}\hat{c}_{ \bm{k}_1}\rangle,
\label{eqn:gamma_gen2}
\end{equation}
The operator $\hat{\gamma}_{ \bm{k}_1,  \bm{k}_2,  \bm{k}_3}= \frac{g}{2}\int \mbox{d}^3r\sum_{ \bm{k}_1,  \bm{k}_2, \bm{k}_3}\phi\left(  \bm{r}\right)e^{\dot{\imath}\left( \bm{k}_1- \bm{k}_2-\bm{k}_3\right)  \bm{r}}$ contains Fourier transform of $\phi\left(  \bm{r} \right)$. We approximate this function by a Gaussian with the width $\sigma$:
\[ \phi\left(  \bm{r} \right) =\left(\frac{1}{2\pi\sigma^2}\right)^{3/4} e^{-\frac{ \bm{r}^2}{4\sigma^2}}\]

Then the operator $\hat{\gamma}$ is simplified to
$ \hat{\gamma}_{ \bm{k}_1,  \bm{k}_2, \bm{k}_3}= \frac{g\left(8\pi\sigma^2\right)^{3/4}}{2}\sum_{ \bm{k}_1,  \bm{k}_2, \bm{k}_3}e^{-\sigma^2\left( \bm{k}_1- \bm{k}_2-\bm{k}_3\right)^2} $

We use the next approximations, typical for derivation of a master equation
\begin{eqnarray*}
\sum_{ \bm{k}} &=& \frac{V}{8\pi^3} \int \mbox{d}^3k \\
\int \mbox{d}^3 k e^{-2\sigma^2\left( {k}_2-( {k}_1-k_3)\right)^2}e^{-\dot{\imath} \tau \frac{\hbar^2  {k}_2^2}{2m}}&=&\left(\frac{\pi}{2\sigma^3}\right)^{3/2}e^{-\dot{\imath} \tau \frac{\hbar^2 \left(( \bm{k}_1-\bm{k}_3)\right)^2}{2m}}\\
\int \mbox{d}\tau e^{\dot{\imath} \tau \hat{\Delta}}&=& \pi \delta\left(\hat{\Delta}\right) + \dot{\imath} P\frac{1}{\hat{\Delta}},
\end{eqnarray*}
where $P\frac{1}{x}$ is a principal value. 
After using these three approximations the expression \eqref{eqn:gamma_gen2} has a form
\begin{equation}
\gamma \hat{a}\hat{\rho}_{BEC}\hat{a}^{\dagger} = -\frac{g^2}{4\hbar^2 \left(2\pi\right)^6} \int \mbox{d}^3  {k}_1 \int \mbox{d}^3 k_3\;\hat{a}\hat{\rho}_{BEC}
\hat{a}^{\dagger} \pi\delta \left(\hat{\Delta}\right) \langle \hat{c}_{ \bm{k}_1}^{\dagger}\hat{c}_{\bm{k}_1}\rangle,
\label{eqn:gamma_gen3}
\end{equation}
In \eqref{eqn:gamma_gen3} we have omitted the term with a principal value. This term should give infinite shift to the energy levels, but does not change the damping coefficient. It appears in many calculations, for example in description of a spontaneous emission -- after the renormalization it represent the Lamb shift. We don't want to discuss this term -- our aim is only a very rough estimation of the damping coefficient.

In the expression \eqref{eqn:gamma_gen3} appears the operator 

\[\hat{\Delta} = \omega_0 +\lambda \hat{a}^{\dagger}\hat{a}-\frac{\hbar^2k_3^2}{2m}+\frac{\hbar}{m}{ \bm{k}_1} \bm{k_3}\]

We assume that background atoms have much higher energy than cold atoms. Furthermore in typical experiments with optical lattices in each lattice site the average number of atoms is small, so the contribution of interaction energy is negligible. Finally with good approximation we have

$\hat{\Delta} \approx -\frac{\hbar^2k_3^2}{2m}+\frac{\hbar}{m}{ \bm{k}_1} \bm{k}_3=\Delta$

Then $\delta\left( -\frac{\hbar^2k_3^2}{2m}+\frac{\hbar}{m} { \bm{k}_1} \bm{k}_3 \right) = - 
\frac{m}{\hbar  {k}_1 {k}_2}\delta\left( cos\theta -\frac{k_3}{ {k}_1}\right) ,$

where $\theta$ is an angle between the wave-vectors ${ {k}_1}$ and ${k_3}$.

After integrating over $k_3$ the equation \eqref{eqn:gamma_gen3} has a form
\begin{eqnarray}
\gamma  = -\frac{g^2\pi m}{4\hbar^3 \left(2\pi\right)^4} \int_0^{\infty} \mbox{d}  {k}_1  {k}_1^3 \langle\hat{c}_{ {k}_1}^{\dagger}\hat{c}_{ {k}_1}\rangle,
\label{eqn:gamma_gen4}
\end{eqnarray}
We calculated the quantity $\langle\hat{c}_{\bm{k}}^{\dagger}\hat{c}_{\bm{k}}\rangle$ from the grand canonical ensemble using the bound $\langle \hat {n}\rangle=\sum_{ \bm{k}}\langle\hat{c}_ {\bm{k}}^{\dagger}\hat{c}_{\bm{k}}\rangle=N, $ where $N$ is the total number of background atoms, and get
\[ \langle\hat{c}_ {k}^{\dagger}\hat{c}_ {k}\rangle =\frac {n}{V} \left(\frac{4\pi \hbar^2}{2mk_B T}\right)^{3/2} e^{-\frac{\hbar^2 k^2}{2mk_B T}}. \]
Finally we can integrating over wave-vectors of incoming background atoms and simplify equation \eqref{eqn:gamma_gen4}:
\begin{equation}
\gamma  = \sqrt{\frac{2}{3}\pi}a^2 \rho_b \langle v \rangle ,
\label{eqn:gamma_gen5}
\end{equation}
where $\rho_b=N/V$ is a density of background atoms and  $\langle v \rangle$ is an average velocity of them. 
The last formula up to a constant can be deduced classically.
\end{widetext}

\end{document}